\documentclass[11pt]{article}

\usepackage[parfill]{parskip}
\usepackage{xcolor}
\usepackage{natbib}
\usepackage{bm}
\usepackage{amsmath}
\usepackage{graphicx}
\usepackage{rotating}
\usepackage{amsfonts}

\usepackage[margin=2.5cm]{geometry}
\usepackage{hyperref} 

\title{Reconciliation of expert priors for quantities and events and application within the probabilistic Delphi method}
\author{Kevin J. Wilson$^{1}$, Malcolm Farrow$^{1}$, Simon French$^{2}$ \& David Hartley$^3$ \\
	$^{1}$School of Mathematics, Statistics \& Physics, Newcastle University, UK \\
	$^{2}$Emeritus, Alliance Manchester Business School, University of Manchester, UK \\
	$^{3}$Walton-On-Thames, Surrey, UK}
\date{}

\begin{document}
	
	\maketitle		
		
\begin{abstract}
We consider the problem of aggregating the judgements of a group of experts to form a single prior distribution representing the judgements of the group. We develop a Bayesian hierarchical model to reconcile the judgements of the group of experts based on elicited quantiles for continuous quantities and probabilities for one-off events. Previous Bayesian reconciliation methods have not been used widely, if at all, in contrast to pooling methods and consensus-based approaches. To address this we embed Bayesian reconciliation within the probabilistic Delphi method. The result is to furnish the outcome of the probabilistic Delphi method with a direct probabilistic interpretation, with the resulting prior representing the judgements of the decision maker. We can use the rationales from the Delphi process to group the experts for the hierarchical modelling. We illustrate the approach with applications to studies evaluating erosion in embankment dams and pump failures in a water pumping station, and assess the properties of the approach using the TU Delft database of expert judgement studies. We see that, even using an off-the-shelf implementation of the approach, it out-performs individual experts, equal weighting of experts and the classical method based on the log score.
\end{abstract}


\section{Introduction}

Typically expert judgement elicitation (expert knowledge elicitation, expert opinion elicitation) to define a subjective prior distribution is conducted with a group of experts. The rationale is that a group, chosen appropriately to cover the range of scientific knowledge on the subject, will provide more reliable and defensible judgements. In most cases we wish to summarise the results of the elicitation as a single prior distribution - for use in Bayesian inference, Monte Carlo simulation, e.g., as part of a sensitivity or uncertainty analysis, or to be provided to a decision maker to inform a decision. While the use the prior is to be put to necessarily affects the interpretation of the resulting distribution, i.e., who the prior should belong to \citep{Fre23}, in all cases the need is for some appropriate summary of the judgements made by the individual experts.

A wide variety of approaches to this aggregation problem have been proposed. Those that have found most widespread application are the classical method, the Sheffield method, the probabilistic Delphi method and the IDEA protocol \citep{EFSA14}. The classical method produces a weighted average of the expert distributions, with weights provided by performance on calibration questions \citep{Qui18}. The Sheffield method produces a consensus distribution via facilitated discussion between the experts \citep{Gos18}. The probabilistic Delphi method uses a combination of anonymised information sharing and multiple rounds of judgements with a final mathematical aggregation \citep{EFSA14}. The IDEA protocol uses a combination of the facilitated discussion from the Sheffield method and mathematical aggregation, most commonly via the classical method \citep{Han18}. The result from all four of these approaches is a probability distribution which does not belong to any real individual, which would be regarded as desirable for a subjective prior in a Bayesian analysis. Only the Sheffield method has attempted to address this - with the experts agreeing that the consensus distribution represents the judgements of a synthetic individual known as the {\em rational impartial observer}. While we are free to adopt the distribution resulting from one of the other methods as representing our beliefs \citep{Fre23}, this would require a clear rationale for the choice of one approach over the others.

Mathematical aggregation via the classical method or the standard use of the IDEA protocol assumes that experts are giving independent judgements on the elicitation quantities. Empirical studies have shown strong correlations between the judgements of experts \citep{Wil17,Wil18} and if, as is true under all formal protocols, all quantitative evidence is shared with all of the experts prior to their judgements, this assumption will always be problematic. In addition, approaches that use calibration questions significantly add to the elicitation burden for the experts and elicitation team, and the choice of suitable calibration questions is a challenging task in itself \citep{Hem21}. Performance-based approaches such as the classical method and IDEA often lead to a very small minority of experts being given weight in the final, aggregated distribution and, in many cases, all weight being given to a single expert.

The Sheffield method, and other consensus-based approaches, can be subject to risks such as dominance of a small number of vocal experts, deference based on social or professional hierarchies and can be prone to cognitive biases such as group-think. In addition, they require all of the experts to participate at the same workshop, whether in person or virtually, and typically require days, rather than hours, to complete if prior distributions on multiple quantities are required. The probabilistic Delphi method, which does not require calibration questions, allows experts to complete the elicitation tasks in their own time and involves limited anonymised interaction between experts through the sharing of judgements and rationales, has some attractive features in this regard. However, the question remains as to how to imbue the probability distribution which results from the probabilistic Delphi method with an interpretation as a subjective prior distribution, and how to capture dependence in the judgements of different experts, particularly those whose judgements are primarily based on the same sources of evidence.

A related area of work is Bayesian reconciliation (also known as Bayesian aggregation). Here we treat the judgements made by the individual experts as observations, represented in a likelihood function. An additional prior distribution is specified, representing an individual external to the experts, typically referred to as the decision maker or investigator. The decision maker's prior is updated via Bayes Theorem to form their posterior distribution. This posterior distribution is the reconciled (or aggregated) prior distribution, and has a direct interpretation as representing the judgements of the decision maker in light of the judgements provided by the experts. The challenge in Bayesian reconciliation is in constructing the likelihood function. Reviews of Bayesian reconciliation are provided in \cite{Fre11,Har21}. While Bayesian reconciliation results in a well-defined prior distribution, it has rarely been used in practice as it has not been embedded in an appropriate elicitation framework. In addition, recently proposed methods have involved the extra elicitation burden associated with eliciting calibration questions. This paper proposes an approach to Bayesian reconciliation which does not rely on calibration questions and embedding it within the probabilistic Delphi method. 

Early approaches to Bayesian reconciliation \citep{Lin83,Win81,Jou96} defined the likelihood function directly. With the advent of modern computing methods for Bayesian inference, hierarchical modelling approaches have been proposed to reconcile expert judgements for continuous quantities \citep{Alb12,Har21}. This allows us to group experts by background, knowledge, experience or rationales behind judgements and incorporate different levels of dependence between the judgements provided by the experts. The judgements of the decision maker are given in the bottom level of the hierarchy. \cite{Alb12,Har21} used parameters constructed from the elicited quantiles in the hierarchical structure, and proposed adjustments to account for the widely observed overconfidence of individual experts \citep{Cle02}. 

In this paper we propose a hierarchical approach to reconcile priors on continuous quantities which combines quantiles specified by the experts directly, while respecting the natural ordering of the quantiles themselves. We extend the hierarchical approach to Bayesian reconciliation to incorporate variables on different scales by first standardising each of the variables using the cumulative distribution function of the decision maker's prior, following the approach of \cite{Wip95}. 

We consider the special case of one-off or non-replicable events, which we will refer to as simply "events". Each of the protocols already discussed could be used to elicit a continuous probability distribution for an event, by considering them as replicable events which take the form of collections of exchangeable Bernoulli trials with underlying rates or proportions. However, in the case of one-off events this is not an appropriate approach, and the elicitation task for an expert reduces to eliciting a single probability for the binary event in question.  All previous hierarchical approaches to Bayesian reconciliation have considered the reconciliation of prior quantiles for continuous quantities. While there have been Bayesian approaches to reconcile expert probabilities for such one-off events \citep{Fre80,Fre81,Cle87}, we propose, for the first time, a hierarchical Bayesian model to reconcile the probabilities provided by the experts for an event. 
As in the case of continuous quantities, the model allows us to define stronger correlations between probabilities given by experts within a group than those between groups.

We give a detailed introduction to the probabilistic Delphi method in Section \ref{sec:delphi}. Section \ref{sec:method} describes the methodology development. We outline our Bayesian hierarchical approach to the reconciliation of expert judgements for continuous quantities, including the standardisation step, in Section \ref{sec:quant} and our hierarchical approach to the reconciliation of event probabilities in Section \ref{sec:probs}. Throughout Section \ref{sec:method} we discuss how the hierarchical approach developed can be embedded within the probabilistic Delphi method. Section \ref{sec:app} provides applications and demonstrates some properties of the approach, focusing on the hierarchical Bayesian models developed. We apply the approach for continuous quantities to a study involving the assessment of erosion in embankment dams by eleven experts in Section \ref{sec:app1} and show general properties by applying the approach to all 45 studies in the TU Delft (Delft University of Technology) expert judgement database in Section \ref{sec:delft}. In Section \ref{sec:app2} we demonstrate the reconciliation approach for event probabilities by applying it to an example assessing whether a pump will fail in a water pumping station during a flash flood. We summarise the paper in Section \ref{sec:sum}.

\section{The probabilistic Delphi method}
\label{sec:delphi}

In this section we provide an outline of the probabilistic Delphi method. For a more comprehensive explanation see Section 6.3 of \cite{EFSA14}. The probabilistic Delphi method shares some common tasks with other formal expert knowledge elicitation approaches, including a careful selection of appropriate experts, training of the experts in the tasks to be conducted, the compilation of an evidence dossier summarising all relevant quantitative information which is shared with the experts and comprehensive reporting post-elicitation. A flow diagram describing the interaction of all of these steps is provided as Figure \ref{fig:delphi} in the Appendix. We suppose all of the preparatory tasks have taken place in the description that follows.

Suppose we have experts $i=1,\ldots,I$ who will undergo the elicitation process for a quantity of interest $\theta$. In the standard approach to the probabilistic Delphi method, three quantiles of expert $i$'s prior distribution for $\theta$, $(q_{1i},q_{2i},q_{3i})$, are elicited, together with lower and upper plausible values for $\theta$, $(\ell_i,u_i)$. This is achieved via a (typically online) survey tool. The Delphi approach uses rounds of elicitation. So in fact the quantities $\bm e_{ij} = (l_{ij},q_{1ij},q_{2ij},q_{3ij},u_{ij})$ are elicited for expert $i=1,\ldots,I$ in round $j=1,\ldots,J$. As standard, two or three rounds will take place. Together with the elicited quantities, a rationale is collected from each expert in each round for the values chosen. 

In rounds $j=2,\ldots,J$ the experts are provided with the anonymised values and rationales provided by all of the other experts, and then asked for their updated quantities. By considering the rationales and interpretations of the other experts, the expectation is that over subsequent rounds the divergence between the values given by different experts will reduce.

Following the final round $J$, the final quantities provided by each expert, $\bm e_{iJ}$, are used to fit a prior distribution with pdf $f_i(\theta)$. We drop the $J$ subscript for brevity. This is typically achieved via least squares using a single parametric form or by comparing the fits of multiple parametric forms. The lower and upper plausible values can be assumed as extreme quantiles (e.g., 0.01 and 0.99 quantiles), or can be excluded from the fitting.  

The standard way to aggregate the $I$ individual expert priors into a single ``consensus'' prior is via an equally weighted linear pool
\begin{displaymath}
	f(\theta) = \sum_{i=1}^I w_i f_i(\theta),
\end{displaymath} 
where $w_i=1/I$. While this intuitively provides an ``average'' of the distributions provided by the experts, and treats each expert equally, it has no clear interpretation as a subjective probability distribution. 

We can see an iterative process in the bottom section of Figure \ref{fig:delphi} in the Appendix. We see that there are two typical stopping criteria: either we stop when no expert changes their judgements or we complete a pre-specified number of rounds. Once the final round is complete, we also need to check that the final result is requisite for the needs of the decision maker. Note that in the diagram we have replaced the equal weights aggregation described above with Bayesian reconciliation, which is the approach we will propose in the next section for the aggregation step.

A similar probabilistic Delphi approach can be used to aggregate probabilities for an event elicited from a group of experts. In this case the experts each provide a probability and the vector of elicited probabilities at stage $j$ is $\bm p_j=(p_{1j},\ldots,p_{Ij})$. Following the same Delphi process of anonymised information sharing and revision of individual probabilities, the final vector of probabilities is given by $\bm p_J$. The equal weights aggregation of these probabilities is $p=1/I\sum_{i=1}^Ip_i$. In this case it is even less clear that an equally weighted average is a sensible choice, and so we will again propose an alternative approach based on Bayesian reconciliation.

\section{Reconciliation of expert priors}
\label{sec:method}

Following the probabilistic Delphi method, the experts will often naturally separate into a small number of groups, based on the rationales they provided for the final quantities they gave. Given the information sharing between rounds of the probabilistic Delphi method, it is not reasonable to assume that judgements given by experts, either within or between these groups, are independent. More generally, when experts are convened to perform elicitations, there are typically subgroups of experts within the larger group, the members of which are likely to have similar beliefs based on similar experience and knowledge, or who use similar reasoning to come up with their values. We may also wish to avoid the dominance, in the aggregated distribution, of subgroups of experts with a larger number of members present.

Therefore, in this section, we propose an approach to the reconciliation of exert priors based on hierarchical modelling, providing specific model structures for the cases of the elicitation of continuous quantities and one-off events. The hierarchical model structure allows us to group experts based on their rationales, and give stronger correlations between elicited quantities from experts within a group compared to experts in different groups. We begin with the reconciliation of continuous quantities, as this is the problem that has been most widely studied in the literature to date. We will discuss the relationship between elicitation for one-off events and continuous quantities once we have discussed Bayesian reconciliation of each.

\subsection{Reconciliation of continuous quantities}
\label{sec:quant}

\subsubsection{The hierarchical model}

Now suppose we are interested in continuous quantities $\bm\theta = (\theta_1\,\ldots,\theta_J)^{'}$ about which we wish to make inference. If the inference is to be Bayesian, we require a prior distribution on $\bm\theta$, which represents our beliefs before we see the data. We suppose that we wish to form an informative prior for $\bm\theta$, and we will ask a group of experts for their beliefs to help us to form this prior. The experts are indexed $i=1,\ldots,I$. More generally, we may never observe data relevant to $\bm\theta$, and so the prior will be a representation of our current state of knowledge about $\bm\theta$.

In the case where the elements of $\bm\theta$ can be assumed to be independent, that is, for the decision maker and each expert no information about $\theta_{j}$ would change their beliefs about $\theta_{k}$ where $k \neq j$, a standard approach to the elicitation task is to ask the experts for quantiles of their distribution of $\theta_j$. All common approaches \citep{Gos18,Han18,Qui18} ask the experts for three quantiles as standard: the median $M_{ij}$, a quantile below the median $L_{ij}$ and a quantile above the median $U_{ij}$, for expert $i$ for $\theta_j$. The latter two quantiles allow us to write down a range of plausible values; $R_{ij}=U_{ij}-L_{ij}$. 

We consider an extra subscript, $g=1,\ldots,G$, representing the subgroup an expert belongs to. Therefore, from expert $i$, in group $g$, for quantity $\theta_j$, we have the triplet $(L_{igj},M_{igj},U_{igj})$. The Bayesian solution to the aggregation problem is to consider the beliefs of an individual, often termed the decision maker. We represent their beliefs before they hear the experts' judgements with a prior distribution, $\pi(\theta_j)$, and treat the quantities elicited from the experts as data. The reconciled distribution is then the decision maker's posterior distribution, which is given by
\begin{equation}
\label{eq:Bayes}
\pi(\theta_j\mid \bm L_j,\bm M_j,\bm U_j) \propto \pi(\theta_j)f(\bm L_j,\bm M_j,\bm U_j\mid\theta_j),
\end{equation}
where $\bm L_j=(L_{11j},\ldots,L_{IGj})^{'}$, $\bm M_j=(M_{11j},\ldots,M_{IGj})^{'}$ and $\bm U_j=(U_{11j},\ldots,U_{IGj})^{'}$ and $f(\bm L_j,\bm M_j,\bm U_j\mid\theta)$ is the likelihood, that is, the decision maker's probability of observing vectors $\bm L_j,\bm M_j,\bm U_j$ given $\theta_j$.

This approach was proposed by \cite{Mor77,Lin83}, and has been widely studied, although not widely applied in practice. Recent approaches to Bayesian reconciliation have utilised hierarchical models \citep{Alb12,Har21}, constructing parameters from the elicited quantiles in the hierarchical structure. We will now define a hierarchical model which combines the quantiles specified by the experts directly.

First we model the decision maker's median using a hierarchical structure. Given that the median is a measure of centrality, we propose a Normal model, taking the form
 \begin{eqnarray}
	M_{igj}\mid\mu_{gj},v_{gj} & \sim & \textrm{N}(\mu_{gj},v_{gj}), \\
	\mu_{gj}\mid \mu_j, \tilde{v}_j & \sim & \textrm{N}(\mu_j,\tilde{v}_j), \\ \label{eq:dmM}
	\mu_j \mid m_{j},\bar{v}_j & \sim & \textrm{N}(m_{j},\bar{v}_j),
\end{eqnarray}
for $j=1,\ldots,J$, where $(\mu_{gj},v_{gj})$ are group-level mean and variance parameters, $(\mu_j, \tilde{v}_j)$ are overall mean and variance parameters and $(m_{j},\bar{v}_j)$ are hyper-parameters to be specified. In addition, we give inverse gamma prior distributions to $(v_{gj},\tilde{v}_j)$, $1/v_{gj}\sim\textrm{Gamma}(a_{gj},b_{gj})$ and $1/\tilde{v}_{j}\sim\textrm{Gamma}(\tilde{a}_{j},\tilde{b}_{j})$ to learn about the within and between group variability respectively. These prior distributions belong to the decision maker. The choice of the values of $(a_{\cdot},b_{\cdot})$ defines the prior correlations between specifications of experts within and between groups. In Section \ref{sec:stand} below we introduce the use of a transformation of the quantiles which renders this use of a Normal distribution appropriate even when the quantity of interest has a bounded range.

We could also give a hierarchical model structure to $L_{igj}$ and $U_{igj}$ directly, but the results would not always respect the relationship $L_{igj}<M_{igj}<U_{igj}$. Instead, we define the differences 
\begin{displaymath}
D_{1igj} = M_{igj} - L_{igj},~D_{2igj} = U_{igj} - M_{igj}.
\end{displaymath}
By defining these two differences, rather than using $R_{ij}$, this allows us to learn about the skewness of expert $i$'s distribution, in addition to the location and spread. We then define a hierarchical model structure on the log-differences of the form
\begin{eqnarray}
	\log(D_{kigj})\mid\delta_{kgj},v_{kgj} & \sim & \textrm{N}(\delta_{kgj},v_{kgj}), \\
	\delta_{kgj}\mid \delta_{kj}, \tilde{v}_{kj} & \sim & \textrm{N}(\delta_{kj},\tilde{v}_{kj}), \\ \label{eq:dmD}
	\delta_{kj} \mid d_{kj},\bar{v}_{kj} & \sim & \textrm{N}(\log[d_{kj}],\bar{v}_{kj}),
\end{eqnarray}
for $k=1,2$, where $(\delta_{kgj},v_{kgj})$ are group-level mean and variance parameters on the log-scale, $(\delta_{kj}, \tilde{v}_{kj})$ are overall mean and variance parameters on the log-scale and $(d_{kj},\bar{v}_{kj})$ are hyper-parameters, with $d_{kj}$ on the original difference scale. As with the medians, we give inverse gamma prior distributions to $(v_{kgj},\tilde{v}_{kj})$ to learn about the within and between group variability respectively. The choice of the hyper-parameters from these distributions controls the within and between group expert correlations. There are alternatives to this log-normal formulation, for example based on gamma and beta distributions. They are beyond the scope of this work.

The aggregated prior distribution then has the quantiles $M_j=\mu_j$ and 
 \begin{eqnarray} \label{eq:L}
	L_j & = & \mu_j - \exp(\delta_{1j}), \\ \label{eq:U}
	U_j & = & \exp(\delta_{2j}) + \mu_j. 	
\end{eqnarray}
We can obtain samples of these quantiles, taken from the posterior distribution given the expert specifications, via MCMC. In this paper, we will use \verb|rjags| \citep{Plu22} for this task. 

There are various parametric and non-parametric forms that could be given to the reconciled distribution based on the three quantiles. A parametric distribution which has the flexibility to represent symmetrical, positively skewed and negatively skewed distributions, and for which the quantile function is available in closed form, is the shifted log-logistic distribution. This has  pdf
$$f(\theta\mid\mu,\sigma,\gamma) = \dfrac{1}{\sigma}\left(1+\dfrac{\gamma(x-\mu)}{\sigma}\right)^{-1/(\gamma+1)}\left[1+\left(1+\dfrac{\gamma(x-\mu)}{\sigma}\right)^{-1/\gamma}\right]^{-2},$$ 
where $\gamma\in\mathbb{R}$ is a shape parameter, $\sigma>0$ is a scale parameter and $\mu\in\mathbb{R}$ is a location parameter. We can express the quantile function $F^{-1}(p)=q$, such that $F(q)=p$, as
\begin{equation}\label{eq:Q}
F^{-1}(p) = \mu + \dfrac{\sigma}{\gamma}\left[\left(\dfrac{p}{1-p}\right)^{\gamma}-1\right].
\end{equation} 
Replacing $F^{-1}(p)$ with $(L_j,M_j,U_j)$ and $p$ with $(p_L,0.5,p_U)$ respectively, and under the condition that $p_L=1-p_U$, i.e., the upper and lower quantiles correspond to probabilities $0.5\pm c$ for some $c<0.5$, some algebra shows that we can express the parameters of the shifted log-logistic distribution as
\begin{eqnarray} \label{eq:alpha}
\mu & = & M, \\ \label{eq:beta}
\sigma & = & \dfrac{\log\left(\dfrac{L-M}{M-U}\right)}{\log\left(\dfrac{p_L}{1-p_L}\right)}\dfrac{ML-UL-M^2+UM}{L-2M+U}, \\ \label{eq:gamma}
\gamma & = & \dfrac{\log\left(\dfrac{L-M}{M-U}\right)}{\log\left(\dfrac{p_L}{1-p_L}\right)}.
\end{eqnarray}

when $U-M\neq M-L$, i.e., the distribution is not symmetrical around the median. When there is such symmetry $\gamma\rightarrow0$ and $\sigma =(L-M)/\log(p_L/(1-p_L))$.

In this way we can use samples of $(L_j,M_j,U_j)$ to obtain samples of $\theta_j$ from the shifted log-logistic distribution. Since $(L_j,M_j,U_j)$ are themselves uncertain, the resulting unconditional reconciled distribution is not itself a shifted log-logistic distribution, only the reconciled distribution conditioned on the three quantiles.

Given the quantiles from the experts, $(L_{igj},M_{igj},U_{igj})$ and the specifications of the hyper-parameters $(a_{\cdot},b_{\cdot})$ for both the median and difference hierarchical models, a summary of the steps to obtain samples from the reconciled distribution is
\begin{enumerate}
	\item Sample $\mu_j^{(k)}, \delta_{1j}^{(k)},\delta_{2j}^{(k)}$, $k=1,\ldots,K$ via MCMC in \verb|rjags|.
	\item Calculate $(L_j^{(k)},M_j^{(k)},U_j^{(k)})$ using $M_j^{(k)}=\mu_j^{(k)}$ and equations (\ref{eq:L}) and (\ref{eq:U}).
	\item Find $(\mu_j^{(k)},\sigma_j^{(k)},\gamma_j^{(k)})$ from $(L_j^{(k)},M_j^{(k)},U_j^{(k)})$ using equations (\ref{eq:alpha}), (\ref{eq:beta}) and (\ref{eq:gamma}).
	\item Sample $\theta_j^{(k)}$ from the shifted log-logistic distribution with parameters $(\mu_j^{(k)},\sigma_j^{(k)},\gamma_j^{(k)})$ by first sampling $u^{(k)}\sim\textrm{Unif}[0,1]$ and then taking $\theta_j^{(k)}=F^{-1}(u_j^{(k)})$ using equation (\ref{eq:Q}).
\end{enumerate}

The shifted log-logistic distribution has the advantage that the expressions for the parameters are available in closed form, which avoids numerical methods for their evaluation. Since the parameters are evaluated for each MCMC sample, this offers a significant computational saving. The shifted log-logistic distribution is also able to incorporate symmetrical, positively skewed and negatively skewed distributions, allowing us to represent a wide range of expert specifications. The shifted log-logistic distribution is bounded when $\gamma\neq0$, that is $\theta_j\geq \mu-\sigma/\gamma$ when $\gamma>0$ (positive skew) and $\theta_j\leq \mu-\sigma/\gamma$ when $\gamma<0$ (negative skew). This will often be a desirable property in fitting distributions to expert judgements. Of course, the method outlined above does not depend of the shifted log-logistic distribution, and other distributions can be used in its place. For example, \cite{Har21} considered the split normal distribution and \cite{Cle02} used two uniform components with exponential tails in their hierarchical models.

\subsubsection{Standardising scales}
\label{sec:stand}

In practice, the quantities of interest $\bm\theta=(\theta_1,\ldots,\theta_J)^{'}$ will often be on different scales, sometimes orders of magnitude different. We prefer to work on the same, standardised, scale for each quantity. As suggested in \cite{Wip95}, we propose to transform the quantiles specified by the experts to $X_{Qigj}=F_j(Q_{igj})$, where $Q_{igj}\in\{L_{igj},M_{igj},U_{igj}\}$ and $F_j(\cdot)$ is the decision maker's prior cdf for $\theta_j$. This transforms the quantiles on to $[0,1]$. We could work with the transformed quantities directly. To utilise the hierarchical model proposed above, we work with the logits of these quantities,
\begin{displaymath}
Z_{Qigj} = \log\left(\dfrac{X_{Qigj}}{1-X_{Qigj}}\right).
\end{displaymath}
Thus, the standardised quantities for the hierarchical model are $\tilde{D}_{1igj}=Z_{Migl}-Z_{Ligj}$, $\tilde{M}_{igj}=Z_{Migj}$ and $\tilde{D}_{2igj}=Z_{Uigj}-Z_{Migj}$. The reconciled quantities from the hierarchical model are $\tilde{\mu}_j,\tilde{\delta}_{1j},\tilde{\delta}_{2j}$ and the standardised reconciled quantiles are found from equations (\ref{eq:L}) and (\ref{eq:U}). The quantiles are back-transformed, first via $$Y_{Qj}=e^{\tilde{Q}_j}/(1+e^{\tilde{Q}_{j}})$$ 
for $Q_{j}\in\{L_{j},M_{j},U_{j}\}$, and then via $Q_j=F^{-1}_j(Y_{Qj})$. 

The decision maker's prior can be elicited directly, if there is a real decision maker for the problem at hand, and this individual has sufficient expertise to provide a prior. If this is not the case, the decision maker's prior can be set to be relatively non-informative over the range of parameter values that are deemed to be plausible. For example, the full range of values that are plausible under the specifications made by the experts in the first round of the probabilistic Delphi method could be used. We could then define the decision maker's prior to be uniform over this range. This would allow for standardised choices to be made for the hyper-parameters in the hierarchical model. This will be discussed in the application in Section \ref{sec:app1}.

\subsubsection{Interpretation of the reconciled distribution}

While hierarchical approaches to the reconciliation of expert judgements such as that outlined in this section, and those given in \cite{Alb12,Har21}, treat the expert elicited quantities as data and update a prior via Bayes Theorem, the decision maker's prior for the median and differences are given in the bottom level of the hierarchy and are related to the individual expert priors in a specific way. Therefore, the interpretation of the resulting aggregated quantities, in the posterior for the bottom level of the hierarchy, needs some care. In \cite{Alb12}, they describe the resulting distribution as {\em  the true parameter ..., or more realistically as the parameter representing the agreement of experts} and similarly \cite{Har21} describe it as {\em the agreement of the experts given the decision maker's prior.}

As a result of the nature of the hierarchical model averaging, in a sophisticated way, the distributions of the experts, the resulting reconciled distribution has the desirable property that the variance of the decision maker's distribution would not tend to zero as the number of experts tends to infinity. Taken together with the parsimony of the approach, we view the hierarchical model as an attractive way to reconcile the judgements of experts. The interpretation we will make is that there is an underlying distribution which represents the population of experts, and we are trying to learn about what it is. Clearly this interpretation will hold as the number of experts becomes large. With a small number of experts, the decision maker's prior will have a greater influence on the reconciled distribution.

\subsection{Reconciliation of events}
\label{sec:probs}

\subsubsection{The hierarchical model}
\label{sec:probs1}

Suppose that we have a number of events of interest $j=1,\ldots,J$, each of which is represented by a binary variable $X_j\in\{0,1\}$. Experts $i=1,\ldots,I$ in groups $g=1,\ldots,G$ will provide their individual assessments of the probability that $X_j=1$, denoted $p_{igj}$ for expert $i$ in group $g$ for event $j$. The task in this case is to obtain a single reconciled probability $p_j$ for each event $j$ to represent the judgement of the decision maker, having heard the probabilities of each of the experts.

That is, we have events $X_j$ such that we are interested in the probability $\Pr(X_j=1)$, $j=1,\ldots,J$. We can represent $X_j$ using a Bernoulli distribution,
\begin{displaymath}
X_j\mid p_j \sim \textrm{Bern}(p_j),
\end{displaymath} 
where $p_j=\Pr(X_j=1)$. We assume that we use the rationales provided by the experts in the final round of the probabilistic Delphi method, or some other appropriate information, to group them into the $G$ groups.

We define a hierarchical model for $p_{igj}$. We suppose that {\em a priori} we believe that the probabilities given by the experts are centred on the true probability, and that the experts' probabilities are correlated, with stronger correlations between experts in the same group, as they used similar rationales in coming to their probabilities. Then we can define
\begin{eqnarray*}
	p_{igj}\mid n_W,p_{gj} & \sim & \textrm{Beta}(n_Wp_{gj},n_W(1-p_{gj})), \\
	p_{gj}\mid n_B,p_j & \sim & \textrm{Beta}(n_Bp_j,n_B(1-p_j)),
\end{eqnarray*}
where $p_{gj}$ is the mean of the beta distribution in the top level, representing the probability of event $j$ in group $g$ and $(n_W,n_B)$ can be thought of as within group and between group prior sample sizes respectively. That is, $\textrm{E}[p_{igj}]=p_{gj}$ and \\ $\textrm{Var}{(p_{igj})=[p_{gj}(1-p_{gj})]/(n_W+1)}$, and similarly for $p_{gj}$. The $(n_W,n_B)$ parameters control the within and between group variances, and the correlations between the probabilities provided by experts in the same group and in different groups respectively.

Rather than specify these prior sample sizes, we give them prior distributions
\begin{eqnarray*}
	n_W\mid a_W,b_W & \sim & \textrm{Gamma}(a_W,b_W), \\
	n_B\mid a_B,b_B & \sim & \textrm{Gamma}(a_B,b_B),
\end{eqnarray*}
and then specify $(a_W,b_W,a_B,b_B)$. This allows us to learn about these parameters from the observations. We will consider how to specify the hyper-parameters in the application in the next section.

We also need to define a prior distribution for $p_j$, which represents the decision maker's beliefs about the probability of the event before observing the probabilities of the experts. One potentially suitable prior, which would represent prior ignoranace of the decision maker, would be $p_j\sim\textrm{Unif}(0,1)$. We will investigate the effect of the choice of this distribution in the application.

Once we observe the probabilities from each expert, we obtain the posterior distribution for $p_j$ and hence the posterior predictive probability for $X_j$. This is our reconciled probability for the event. That is, the quantity of interest is
\begin{eqnarray*}
	\Pr(X_j=1\mid p_{11j},\ldots,p_{IGj}) & = & \int p_j\pi(p_j\mid p_{11j},\ldots,p_{IGj})dp_j \\
	& \approx & \dfrac{1}{K}\sum_{k=1}^Kp_j^{(k)},	
\end{eqnarray*}
where $p_j^{(k)}$ are samples from the posterior distribution of $p_j$, given by $\pi(p_j\mid p_{11j},\ldots,p_{IGj})$. We can obtain the reconciled probability as above or by drawing $X^{(k)}$ and recording the proportion of occasions in which $X_j^{(k)}=1$. In both cases, we can use MCMC in \verb|rjags| for this task.

\subsubsection{One-off events or proportions}
\label{sec:prop}

In Section \ref{sec:probs1} we have considered a binary variable $X$ which is equal to 1 if an event occurs and 0 if not. In this case the probability distribution for $X$ is captured by a single probability, $\Pr(X=1)$, with $\Pr(X=0)=1-\Pr(X=1)$. This necessarily captures all of our uncertainty on whether the event will occur, both our epistemic (lack of knowledge) uncertainty about the underlying proportion of similar occasions on which the event would occur, and our aleatory (random variability given perfect knowledge) uncertainty about what would happen on this particular occasion.

If we are able to consider an appropriate population of events which we are willing to assume are exchangeable with the event of interest, then we could consider these two aspects separately. That is, we could ask experts to consider a large number of exchangeable situations, and ask for their beliefs about the proportion of occasions on which the event would occur, which we call $\theta$. Questions about quantiles of $\theta$, for example, would allow us to fit a suitable continuous probability distribution for each expert and apply a hierarchical model for reconciling quantiles such as the model proposed in Section \ref{sec:quant}. We would then obtain $Pr(X=1)$ as the posterior predictive probability
\[
Pr(X=1)=E[\theta]=\int_{0}^{1}\theta\times\pi(\theta\mid\bm L,\bm M,\bm U)d\theta,
\] 
where $\pi(\theta\mid\bm L,\bm M,\bm U)$ is the posterior distribution of $\theta$ given expert-specified quantiles $\bm L=(L_1,\ldots,L_I)^{'}$, $\bm M=(M_1,\ldots,M_I)^{'}$ and $\bm U=(U_1,\ldots,U_I)^{'}$ for $\theta$.
However, for one-off events it will often not be sensible to ask the experts about $\theta$, as a suitable population of exchangeable events will not be appropriate, even as a thought experiment, and so instead we assume that we ask them directly for $\Pr(X=1)$.

\section{Applications}
\label{sec:app}

In each of the applications in this section we consider elicitations which have been carried out previously, and for which the quantities elicited from each of the experts individually are available. We will use the applications to assess the hierarchical models for Bayesian reconciliation we proposed in the previous section. While we do not have expert elicited data available from the probabilistic Delphi method to illustrate the full approach to elicitation and reconciliation proposed in this paper, we will highlight the impact that individual elicitations conducted using the probabilistic Delphi method could have had within the applications where appropriate.

\subsection{Application 1: Continuous quantities}
\label{sec:app1}

We consider a study from \cite{Bro04} looking into a cost effective means of early detection of progressive internal erosion in embankment dams. The aim of the study was to aid management of erosion. The unknowns of interest, on which there were no data available at the time the decision support was required, were the prevalence of leakage and internal erosion, the average leakage and erosion rates, the minimum detectable leakage rate and dam critical flow and the rate of deterioration of the dams.

There were 11 experts consulted in the elicitation: seven engineering consultants, two working utility engineers and two academics. The original elicitation was conducted using the classical method, and so 11 calibration questions were asked prior to the elicitation of the study variables. Calibration variables are quantities whose true values are known to the elicitation team but not the experts, and are used in the classical method to define the weights given to the judgements of each expert. It is these 11 calibration variables that we will concentrate on in this example. 

The elicitations took place during two workshops, with technical discussions between the experts taking place prior to the individual elicitations. With multiple rounds of judgements, interaction between the experts and individual judgements, the elicitation exercise contained some of the features of the probabilistic Delphi method. Given the discussion between the experts, it is not plausible that the experts would be providing independent judgements, and so a key assumption of the classical method is not satisfied. We will demonstrate how we might model these dependencies between experts using Bayesian reconciliation. 

The groupings of experts we use are by type of expert: consultants, working engineers and academics - as experts with similar backgrounds are likely to give more correlated responses than those with different backgrounds. If the rationales from the experts were available, then we could use these rationales to form the groupings as described earlier. Full details of the study, and reports of the elicitations carried out, are provided in \cite{Bro04}.

For each of the 11 calibration questions, the experts provided their 5\%, 50\% and 95\% quantiles. That is, the experts specified $(L_{igj},M_{igj},U_{igj})$ such that $\Pr(\theta_j<L_{igj})=0.05$, $\Pr(\theta_j<M_{igj})=0.5$ and $\Pr(\theta_j<U_{igj})=0.95$. We will fit the hierarchical Bayesian model from Section \ref{sec:quant} to reconcile the experts' judgements. We include the standardisation of the judgements as detailed in Section \ref{sec:stand}. For the Bayesian hierarchical model we define the decision maker's prior to be uniform over the full range of plausible values as defined by the experts and we choose the model hyper-parameters to be $a=2,b=2$ for each of the gamma priors on the precision parameters, a prior mean and variance of $m_j=0,\bar{v}_j=100$ on the overall median and prior means and variances on the log distances of $\log(d_{kj})=0,\bar{v}_{kj}=100$.

\subsubsection{Results}

\begin{figure}[tb]
	\centering
	\includegraphics[width=0.49\linewidth]{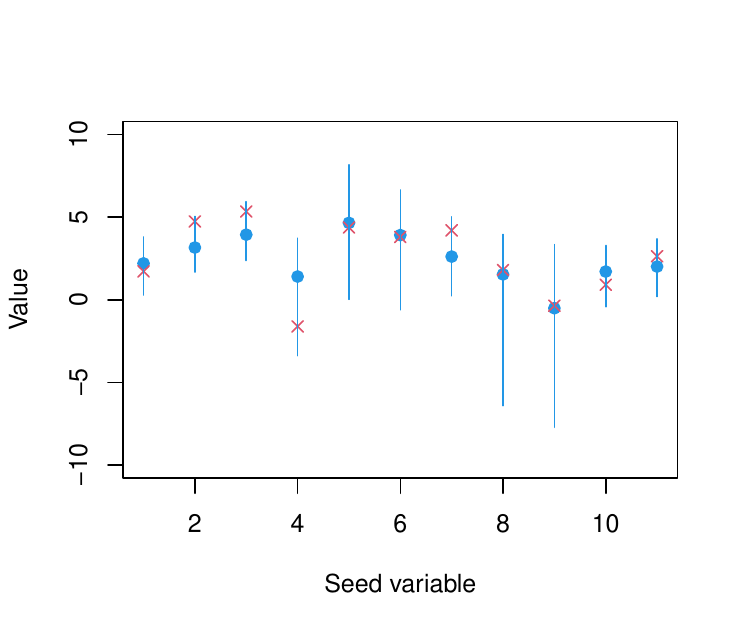}
	\includegraphics[width=0.49\linewidth]{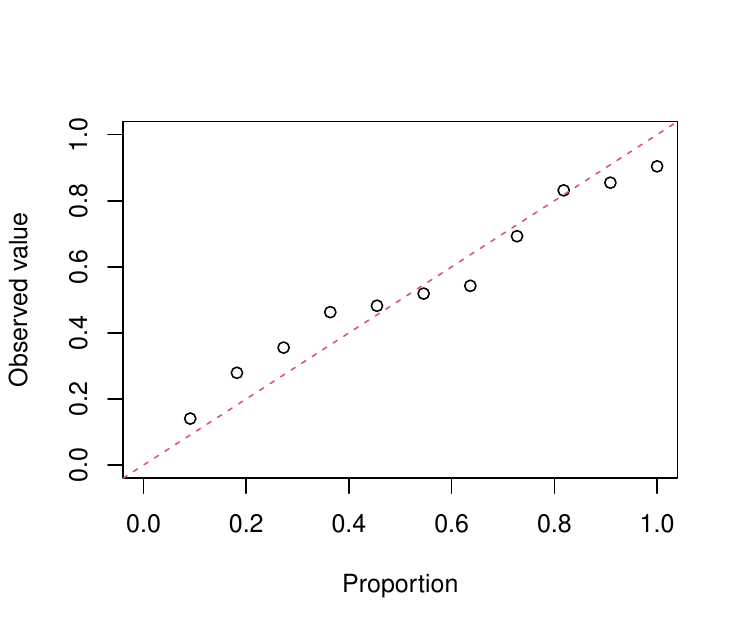}
	\caption{Each calibration variable value (red crosses) and the corresponding median and 90\% symmetric predictive interval (blue) for the reconciled distribution belonging to the decision maker (left) and the empirical cdf value for each of the calibration variables within the reconciled distribution plotted against the proportion under a perfectly calibrated model (right). The red dashed line is at $x=y$.}
	\label{fig:all}
\end{figure}

For each of the 11 calibration questions, we fit the hierarchical Bayesian reconciliation, and then plot the median and symmetric 90\% predictive interval of the reconciled distribution. This is given in the left hand side of Figure \ref{fig:all}. We also plot the empirical cdf value for each of the calibration variables within the reconciled distribution against the proportion under a perfectly calibrated model. That is, we order the 11 calibration values by their quantile position in the reconciled distribution, and then plot this against the values $(1/11, 2/11,\ldots)$. If the reconciled distributions were perfectly calibrated, the points would all lie on the dashed $x=y$ line. This is provided in the right hand side of Figure \ref{fig:all}.

We see that the true value of each calibration variable lies within the body of the reconciled distribution. We can see no clear bias from the method for the 11 calibration variables, with the true values lying both above and below the medians of the reconciled distributions. Similarly, we see realisations in the central portion and both tails of the reconciled distributions. This is confirmed with an overall well-calibrated distribution of values seen in the right hand plot.

\begin{sidewaysfigure}[h!]
	\centering
	\includegraphics[width=\linewidth]{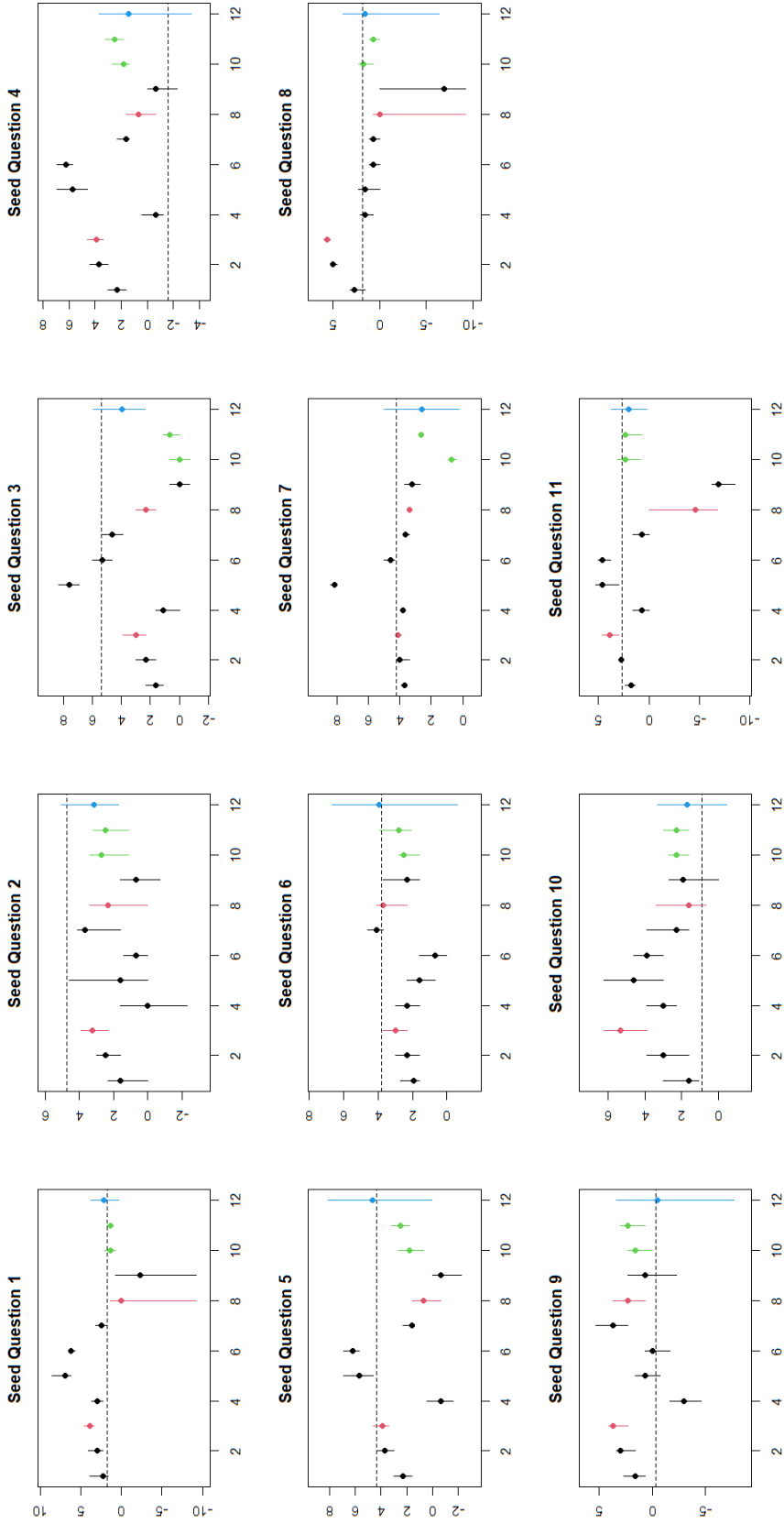}
	\caption{The medians and symmetric 90\% predictive intervals for each expert and the reconciled distribution (blue), for all 11 calibration questions. The experts are coloured by group: group 1 (engineering consultants) in black, group 2 (working utility engineers) in red and group 3 (academics) in green. The dashed line represents the true value of the calibration variable.}
	\label{fig:exp}
\end{sidewaysfigure}

We can also consider how the aggregated distributions relate to the quantiles specified by the individual experts. In Figure \ref{fig:exp} we provide the medians and symmetric 90\% predictive intervals for each expert, coloured by their group, and the reconciled distribution for the 11 calibration questions. In each plot the dashed line represents the true value of the calibration variable.

\clearpage

We see that individually the experts tend not to include the true value of the calibration variables within the body of their distributions the majority of the time, with all of the experts displaying a degree of over-confidence (intervals that are too narrow). This is consistent with previous findings \cite{Spe10,Wil17} We also see some increased homogeneity between experts within groups, particularly the two academics. The reconciled distribution, giving weight to each group of experts, tends to produce much wider intervals, which are then much better calibrated. We note the unusual lower tail in the reconciled distribution for calibration variable 9. We hypothesise this is due to the shifted log-logistic distribution fitting a distribution which is bounded above in this case.

We can assess the predictive abilities of the individual experts and the reconciled distribution using proper scoring rules for probabilistic forecasts. We use the average log-score across the calibration variables. This is given by
\begin{displaymath}
S(\bm\theta) = \dfrac{1}{J}\sum_{j=1}^J-\log\left[f_j(\theta_j)\right],
\end{displaymath}
where $f_j$ is the expert or reconciled density function for calibration variable $j$, and $\theta_j$ is the observed value of the calibration variable. The log-score is an attractive proper scoring rule in this context since it is the only proper scoring rule that depends only on the value of the function at the calibration variable value \citep{Oha06}. A higher score indicates a better predictive distribution. To calculate the log-score for each expert, we fit a shifted log-logistic distribution to their three quantiles. We also compare to an equally weighted aggregation of the experts' distributions. In this case, the aggregated prior distribution is
\begin{equation}
\label{eq:lp}
f_j(\theta_j) = \dfrac{1}{I}\sum_{i=1}^{I}\pi_i(\theta_j),
\end{equation}
where $\pi_i(\theta_j)$ is the  shifted log-logistic distribution for expert $i$ for quantity $j$.

The log-scores for each of the 11 experts, the equally weighted aggregation and the reconciled distribution are given in Table \ref{tab:score}.

\begin{table}
\centering
\begin{tabular}{|ccc|ccc|} \hline
Expert & Group & Score & Expert & Group & Score \\ \hline
1 & 1 & -5.16 & 7 & 1 & -8.14 \\
2 & 1 & -4.33 & 8 & 2 & -6.41 \\
3 & 2 & -4.55 & 9 & 1 & -4.69 \\
4 & 1 &  -4.26 & 10 & 3 & -2.78 \\
5 & 1 & -3.10 & 11 & 3 & -5.97 \\
6 & 1 & -6.25 & && \\  \hline
\end{tabular}

\vspace{4pt}

\begin{tabular}{|cc|} \hline
Method &  Score  \\ \hline
Equal weights &  -1.70  \\
Reconciled &  -0.68  \\ \hline
\end{tabular}
\caption{The log-scores of each of the 11 experts, an equally weighted distribution and the reconciled distribution. A higher score indicates better predictions.}
\label{tab:score}
\end{table}

We see that expert 10 is the best individual expert based on the log-scores and expert 7 is the poorest. No expert has a log-score close to that of the reconciled distribution or the equally weighted distribution. It is clearly superior to reconcile the individual experts' distributions in this case, based on the log-score. The reconciled distribution using the Bayesian hierarchical model also has a higher log-score than the equally weighted distribution.

\subsubsection{TU Delft database of expert judgement studies}
\label{sec:delft}

In \cite{Coo07}, the authors explored a database of 45 expert judgement studies conducted by TU Delft. In all, the studies involved over 7000 expert elicited probability distributions. The studies were conducted in a range of sectors, for substantive problems, including the nuclear industry, the chemical and gas industry, the aerospace sector, health, banking, volcanoes and dams. In each study a number of calibration variables were elicited. The experts in each study were asked for their median and 5\% and 95\% quantiles for each variable, though in a few cases other quantiles were elicited in addition. The same quantiles were elicited for the quantities of interest in the study. For further information on the TU Delft database see \cite{Coo07,Wil17}.

Here we use the approach proposed in Section \ref{sec:quant} to reconcile the judgements provided by the experts in each of the 45 studies. We include the standardisation step in the reconciliation in each case. We compare the resulting reconciled distribution, in terms of the log-score, with a randomly selected expert in each study, the equally weighted distribution and the optimal decision maker using the classical method. The classical method uses the weighted linear pool given in Equation (\ref{eq:lp}), with weights calculated based on the performance of the experts on the calibration questions. The weight for an expert is proportional to their overall score, which is the product of a statistical accuracy score and an informativenss score. The subset of experts which maximises the linear pool's overall score is used as the aggregated distribution. For full details see \cite{Qui18}.

\begin{figure}
\centering
\includegraphics[width=0.49\linewidth]{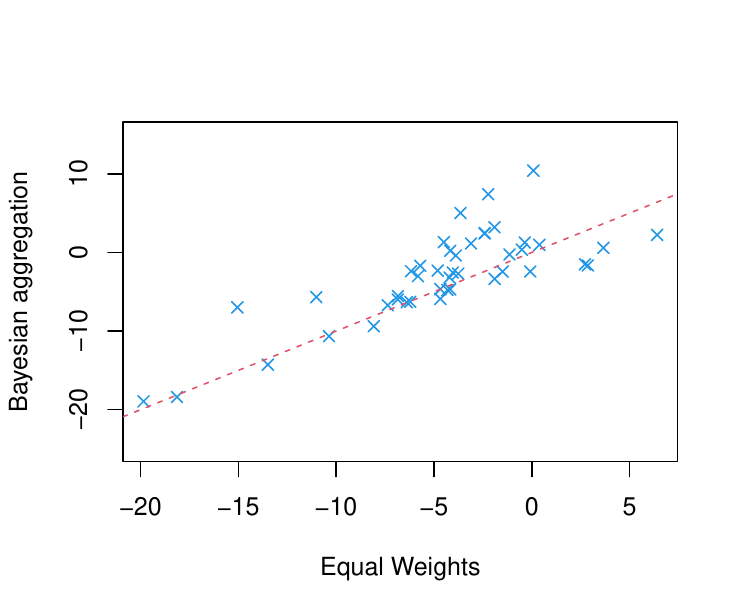}
\includegraphics[width=0.49\linewidth]{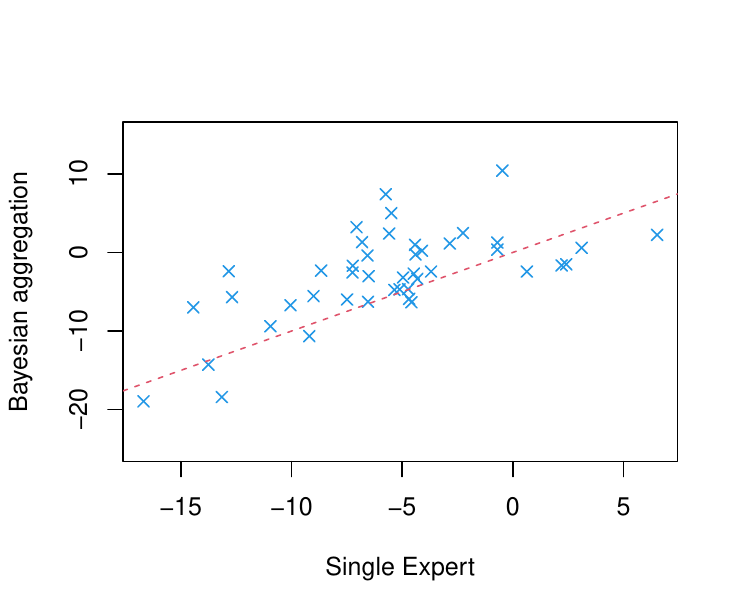}
\includegraphics[width=0.49\linewidth]{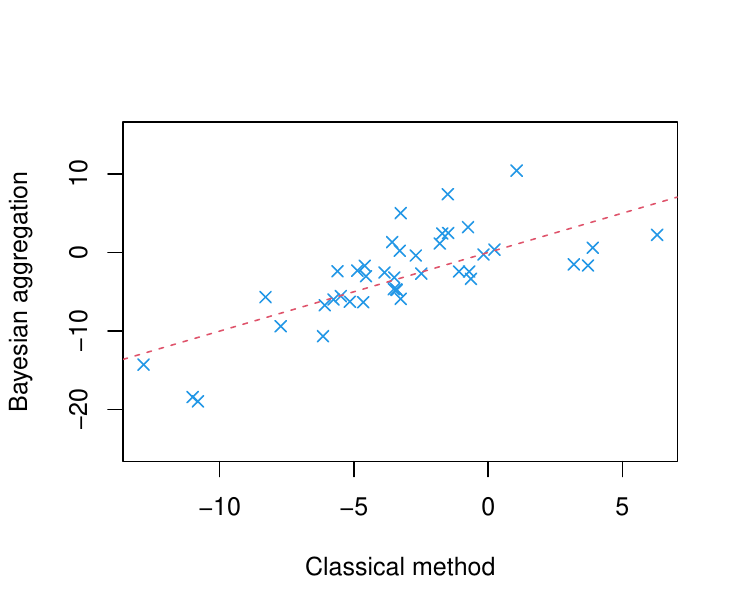}
\caption{The log-scores of an equally weighted distribution (top left), a randomly selected expert (top right) and the optimal decision maker from the classical method (bottom), each against the reconciled Bayesian distribution, for the 45 studies in the TU Delft expert judgement database \citep{Coo07}.}
\label{fig:score}
\end{figure}

Due to the off-the-shelf nature of the comparison, no grouping of experts will be used in the comparison. Therefore, it is sensible to think of the results for the Bayesian reconciliation as a conservative estimate of the true performance of the approach. For the classical method, a leave-one-out approach is used in which the aggregated distribution for a calibration variable uses the optimal linear pool based on all of the other calibration variables.  The results are provided in Figure \ref{fig:score}.

We see that in all three cases the reconciled distribution produces a higher log-score on the majority of occasions, and this result is strongest when comparing to individual experts and weakest comparing to the classical model. Under a Normal model for the difference in the log-scores, with a diffuse prior centred at zero and a diffuse inverse-gamma prior on the variance, the posterior probability that the reconciled distribution has a higher mean log-score than equal weights, given these 45 studies, is 0.9990, the posterior probability that the reconciled distribution has a higher mean log-score than an individual expert is 0.9999 and the posterior probability that the reconciled distribution has a higher mean log-score than the optimal decision maker from the classical method is 0.6607.

\subsection{Application 2: One-off events}
\label{sec:app2}

We consider a simulated set of probabilities, based on an elicitation conducted as part of the UK Climate Resilience Demonstrator (CReDo),  a climate change adaptation digital twin project looking at the impact of flooding on energy, water and telecoms networks. Within CReDo, the assesment of individual asset failure given flooding at a particular location was evaluated using Bayesian networks, with the conditional probability tables populated using probabilities elicited from experts from the relevant asset network \citep{Den22}. We consider one such set of synthetic probabilities for the event ``pump failure'' within a water pumping station. The elicitation was conducted in a workshop, during which the experts gave individual probabilities for the event of interest. The rationales of each of the individual experts were available. In scaling up the elicitations to multiple probabilities for multiple assets over multiple future flooding scenarios, in future iterations the probabilistic Delphi method will be used to elicit the individual probabilities.

The probabilities for the event were elicited from 10 experts. Based on their rationales, they were separated into two groups. The individual probabilities and group membership are provided in the top section of Table \ref{tab:probs}. We see that individuals in group 1 gave the event systematically higher probabilities than experts in group 2.

\begin{table}[tb]
	\centering
	\begin{tabular}{|ccc|ccc|}\hline
		Expert & Group & Probability & Expert & group & Probability \\ \hline
		1 & 1 & 0.44 & 6 & 1 & 0.35 \\
		2 & 1 & 0.49 & 7 & 1 & 0.41 \\
		3 & 1 & 0.48 & 8 & 1 & 0.41 \\ 
		4 & 1 & 0.39 & 9 & 2 & 0.09 \\ 
		5 & 1 & 0.33 & 10 & 2 & 0.11 \\ \hline
	\end{tabular}
	
	\vspace{4pt}
	
	\begin{tabular}{|cc|} \hline
		Method & Probability \\ \hline
		Equal weights & 0.35 \\
		Group 1 & 0.41 \\
		Group 2 & 0.10 \\
		Reconciled & 0.30 \\ \hline
	\end{tabular}
	
	\caption{Top: the individual probabilities from the 10 experts for the event ``pump failure''. Bottom: the equally weighted, group 1, group 2 and reconciled probabilities.}
	\label{tab:probs}
\end{table}

We use the Bayesian hierarchical model from Section \ref{sec:probs}. We choose the values of the hyper-parameters to be $a_W=a_B=20, b_W=b_B=2$, representing prior sample sizes of approximately 10 within and between groups. The resulting reconciled probability, along with the equally weighted average and the mean probabilities from groups 1 and 2, is given in the bottom section of Table \ref{tab:probs}. We see that the reconciled probability is lower than that when applying equal weights to each expert - this seems to be a result of the higher correlations between probabilities from experts in the same group resulting in a more equal weighting of the probabilities in the two groups.

\begin{figure}[tb]
	\centering
	\includegraphics[width=0.49\linewidth]{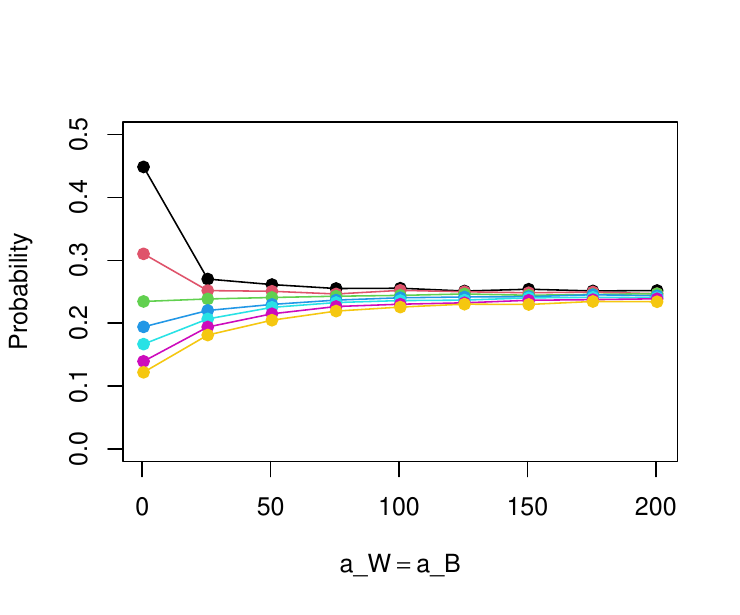}
	\includegraphics[width=0.49\linewidth]{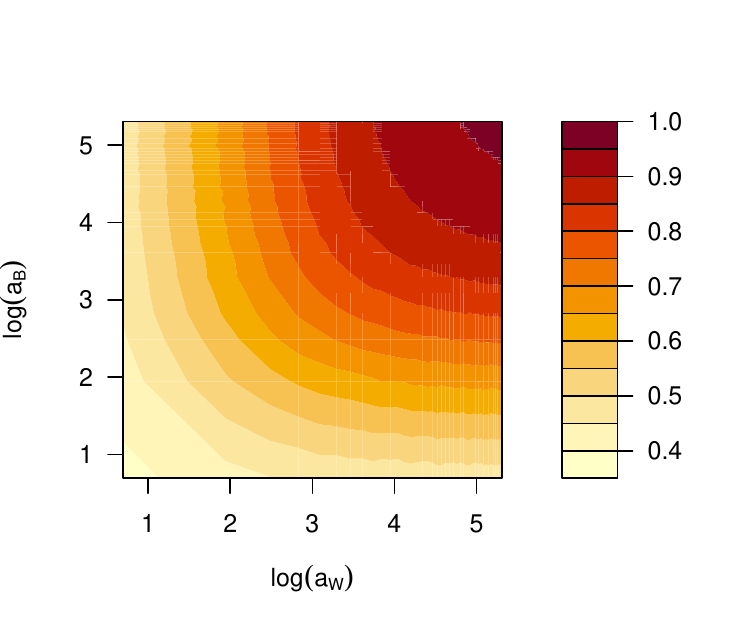}
	\caption{Left: The reconciled probability based on the 10 individual expert probabilities, for values of $a_W=a_B=2,\ldots,200$ and $b_W=b_B=2$. From top to bottom the lines represent decision maker prior means of 0.5 to 0.05. Right: the prior correlation between the probabilities provided by experts in different groups, based on $a_W=a_B=2,\ldots,200$ (on the log scale) and $b_W=b_B=2$.}
	\label{fig:probs}
\end{figure}

We can investigate the effect of the choice of the prior hyperparameters. In the left hand side of Figure \ref{fig:probs} we plot the reconciled probability based on a range of choices for the hyperparameters and decision maker's prior. Specifically we set $a_W=a_B=2,\ldots,200$ and keep $b_W=b_B=2$. The individual lines are based on decision maker's priors in the form of beta distributions with (from top to bottom) equally spaced prior means from 0.5 to 0.05.  

We see that the reconciled probability is relatively close to the decision maker's prior mean for small prior sample sizes (i.e., small values of $a_W$ and $a_B$). As the prior sample size increases, more weight is put on the probabilities provided by the experts, and the reconciled probability becomes very similar no matter what the decison maker's original prior. In the right hand side of Figure \ref{fig:probs} we plot the prior correlations between the probabilities from two experts in different groups, based on different combinations of prior hyperparameter values on the log scale. We see that, as either $a_B$ or $a_W$ increases, i.e., as the prior sample sizes increase, the prior correlation between the probabilities increases relatively quickly in this model.

\section{Summary}
\label{sec:sum}

In this paper we have considered the problem of combining the judgements of a group of experts to form a single prior distribution for continuous quantities and one-off events. The approach we have taken is via Bayesian reconciliation, proposing a novel hierarchical model to reconcile expert judgements for continuous quantities which standardises the variables to allow the reconciliation of quantities on different scales and performs the reconciliation directly on the standardised versions of the elicited quantities. In the case of probabilities for events, ours is the first hierarchical approach we are aware of that performs Bayesian reconciliation. 

To encourage uptake of the methods developed in this paper we have embedded them within the probabilistic Delphi method. This is a mature elicitation protocol which embeds the elicitation principles of rigour, reproducibility and transparency. We believe that an elicitation conducted using the probabilistic Delphi method in which the judgements following the final round of elicitations are reconciled using the Bayesian approach outlined would satisfy each of these properties and also retain an interpretation as a subjective probability distribution - belonging to the decision maker. We name the resulting approach the Bayesian Delphi method.

There are various promising areas for further work. The first is to build up experience of the approach by applying it prospectively to expert judgement studies. It is only in this way that we will be able to fully evaluate the utility of the method. There are several methodological changes which can be made to the hierarchical structures - increasing the flexibility of the model and allowing for further borrowing of information - for example the borrowing of information across elicitation quantities within an expert. Another possible direction would be in the recalibration of expert quantiles as proposed by \cite{Cle02,Har21}.     

\section*{Acknowledgements}

We would like to thank Willy Aspinall for discussions about, and his hard work on, the dam application and Roger Cooke for making the data from the Delft expert judgement available to us. We would also like to thank Jim Smith and Chris Dent for their work on the elicitation aspects of the CReDo project.

\bibliography{refs}
\bibliographystyle{plainnat}

\newpage

\section*{Appendix}

\vspace{-0.5cm}

\begin{figure}[ht]
	\centering
	\includegraphics[width=0.91\linewidth]{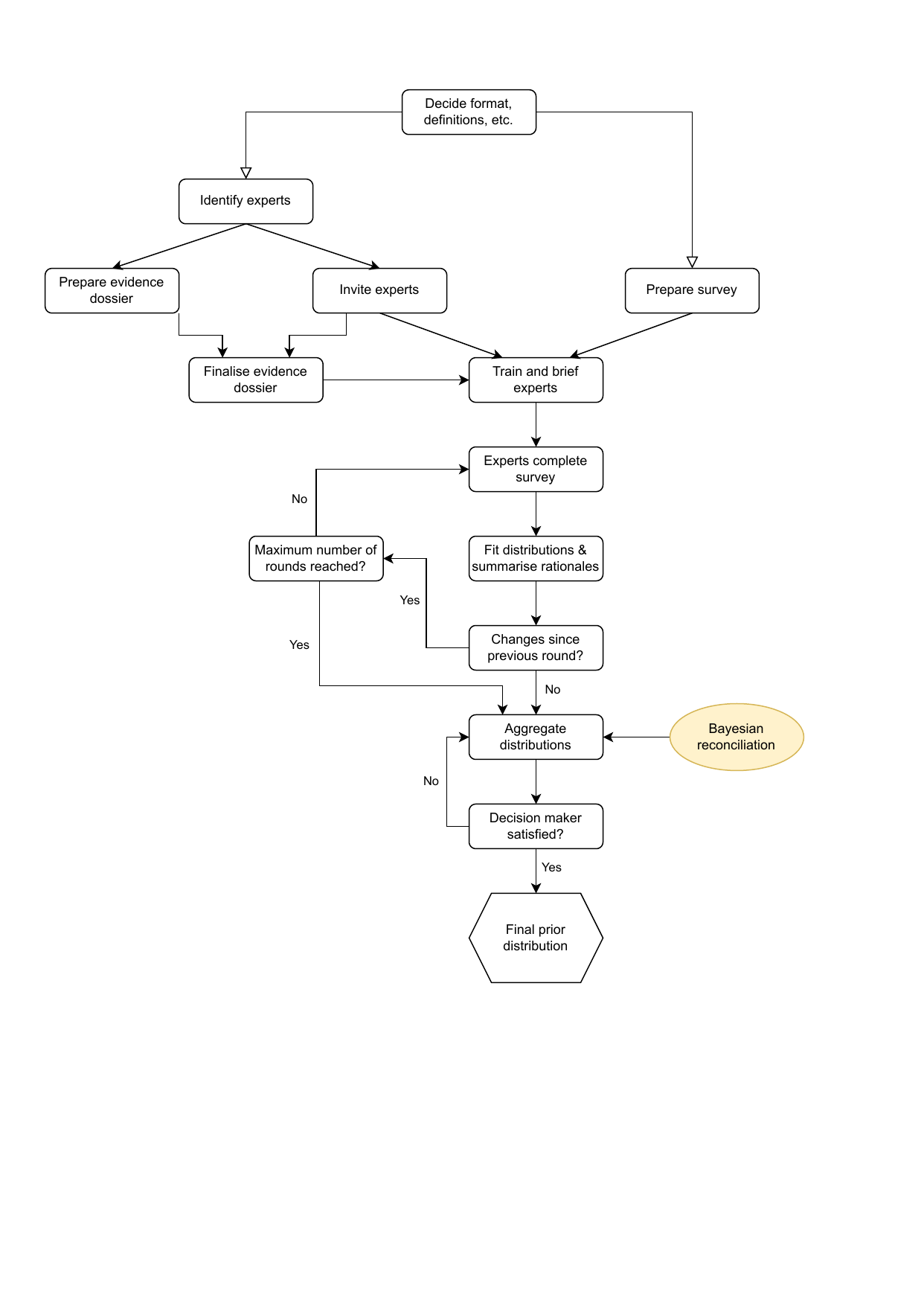}
	\vspace{-4.9cm}
	\caption{The main steps of the probabilistic Delphi method for structured expert judgement elicitation, with Bayesian reconciliation in place of equal weights aggregation.}
	\label{fig:delphi}
\end{figure}

\end{document}